\documentclass[%
preprint,
superscriptaddress,
 amsmath,
 amssymb,
 aps,
]{revtex4-2}

\usepackage{graphicx}% Include figure files
\usepackage{dcolumn}% Align table columns on decimal point
\usepackage{bm}% bold math
\usepackage{placeins}
\begin{document}

\title{Mode transition ($\alpha-\gamma$) and hysteresis in microwave-driven low-temperature plasmas}% Force line breaks with \\

\author{Kyungtae Kim}
\affiliation{Division of advanced nuclear engineering, Pohang University of Science and Technology (POSTECH)}

\author{Woojin Nam}
\affiliation{Department of physics, Pohang University of Science and Technology (POSTECH)}

\author{Jimo Lee}
\affiliation{Division of advanced nuclear engineering, Pohang University of Science and Technology (POSTECH)}
\affiliation{Mechatronics Research, Samsung Electronics Co., Ltd.}

\author{Seungbo Shim}
\affiliation{Mechatronics Research, Samsung Electronics Co., Ltd.}

\author{Gunsu S. Yun}
\email{gunsu@postech.ac.kr}
\affiliation{Division of advanced nuclear engineering, Pohang University of Science and Technology (POSTECH)}
\affiliation{Department of physics, Pohang University of Science and Technology (POSTECH)}

%%%%%%%%%%%%%%%%%%%%%%%%%%%%%%%%%%%%%%%%%%%%%%%%%%%%%%%%%%%%%%%%%%%%%%%%%%%%%%%%%%%%%%%%%%%%%%%%%%%%%%                                               Abstract
%%%%%%%%%%%%%%%%%%%%%%%%%%%%%%%%%%%%%%%%%%%%%%%%%%%%%%%%%%%%%%%%%%%%%%%%%%%%%%%%%%%%%%%%%%%%%%%%%%%%%%

\begin{abstract}
We discovered a hysteresis in a microwave-driven low-pressure argon plasma during gas pressure change across the transition region between $\alpha$ and $\gamma$ discharge modes. The hysteresis is manifested in that the critical pressure of mode transition depends on the direction of pressure change. As a corollary, the plasma would attain different discharge properties under the same operating parameters (pressure, power, and gas composition), suggesting a bi-stability or existence of memory effect. Analysis of the rotational and vibrational temperatures measured from the OH (A-X) line emissions shows that the hysteresis is mainly due to the fast gas heating in the $\gamma$-mode leading to a smaller neutral density than that of the $\alpha$-mode. When increasing the gas pressure, the $\gamma$-mode discharge maintains a relatively higher temperature and lower neutral density, and thus, it requires a higher operating pressure to reach the $\alpha$-mode. On the other hand, decreasing the pressure while maintaining $\alpha$-mode, the transition to $\gamma$-mode occurs at a lower pressure than the former case due to a relatively higher neutral density of $\alpha$-mode discharge. This interpretation is supported by the fact that the hysteresis disappears when the plasma properties are presented with respect to the neutral gas density instead of pressure.
\end{abstract}

\maketitle

\pagebreak

%%%%%%%%%%%%%%%%%%%%%%%%%%%%%%%%%%%%%%%%%%%%%%%%%%%%%%%%%%%%%%%%%%%%%%%%%%%%%%%%%%%%%%%%%%%%%%%%%%%%%%                                               Introduction
%%%%%%%%%%%%%%%%%%%%%%%%%%%%%%%%%%%%%%%%%%%%%%%%%%%%%%%%%%%%%%%%%%%%%%%%%%%%%%%%%%%%%%%%%%%%%%%%%%%%%%

\section{Introduction}
\label{sec:intro}
Radio-frequency (RF) sources around 10 MHz have been widely used to generate plasma discharges, owing to relatively low cost and low technical barrier \cite{MM}. Recently, with advancement in the microwave technologies, plasma sources with higher operating frequencies about 1 GHz have rapidly attracted interest from various plasma applications, such as biomaterials synthesis, polymer surface functionalization, and semiconductor fabrication \cite{musil1986microwave,vesel2017new,forch2005soft}, as they can produce higher electron density and higher population of energetic electrons with comparatively low power \cite{MM, lebedev2015microwave}. In particular, microwave-driven plasmas are successfully applied to the semiconductor manufacturing processes including deposition, etching, and ashing, by varying operating parameters diversely \cite{donnelly2013plasma,bera2008reliability1,bera2008reliability2}. Since these processes are repeatedly applied in high-precision semiconductor fabrication, it is crucial to ensure that the same plasma properties are reproduced under the same process parameters \cite{repeatedly, processingsequence}.

Ideally, an identical plasma state would be reproduced when the external operation parameters are set to the original values as long as the discharge mode is preserved \cite{fu2020similarity1}. However, the situation becomes complicated when the operation parameter space of a plasma process spans across $\alpha$ and $\gamma$ discharge modes. Even under nominally identical operation parameters, the plasma may manifest different properties in electron and ion densities and temperatures since the electron loss and generation mechanisms change significantly crossing the discharge modes. 

The discharge mechanisms of $\alpha$- and $\gamma$-modes can be explained with the ionization by electron-neutral collisions and the secondary electron generation by the bombardment of fast ions to the electrode, respectively \cite{godyak1992evolution,lee2017extended,lee2017scalings,shi2006evolution}. The transition between the two modes is intuitively understood by comparing the collective oscillation amplitude of electrons and the system dimension (the electrode gap distance)  \cite{lee2017extended,lee2017scalings,lee2018generation}.
The $\gamma$ mode corresponds to the situation of the oscillation amplitude being larger than the gap distance.
Interestingly, particle-in-cell simulations~\cite{lee2017extended} of microwave-driven atmospheric pressure plasmas showed a 'Z'-shaped frequency dependence of the gas breakdown voltage, i.e., three different breakdown voltages at the same driving frequency in the mode transition region. That observation alluded a possible existence of hysteresis when a plasma process runs in the vicinity of mode transition. 

This paper reports an experimental study on low-pressure microwave-driven argon plasmas in the vicinity of mode transition, where a hysteresis is found for the change of gas pressure.
Our study proposes that the nonlinear behavior of the plasma occurs due to the fast gas heating effect in $\gamma$-mode where ions are accelerated in the sheath region. This conjecture is supported by a sharp increase of rotational temperature in $\gamma$-mode, reflecting the increase of the translational temperature \cite{lavrov1978relation}. Furthermore, this interpretation suggests that the hysteresis may disappear by replacing the abscissa from gas pressure to neutral density, which is experimentally confirmed. 
It is worth noting that the hysteresis in the microwave-driven plasmas is different from the well-known hysteresis observed in the inductively coupled RF plasmas during the E-H mode transition \cite{LeeHC1, LeeHC2}.

Our work shows that there exist bi-stable plasma states across the mode transition regime depending on the direction of pressure change. Such memory-like property of plasma in the vicinity of mode transition has an immediate implication on high-precision plasma processes where reproducible plasma states are required for high yield. 
A countervailing technique for the hysteresis is proposed to ensure the reproducibility of plasma state, i.e., to control the neutral density instead of the gas pressure.

%\pagebreak

%%%%%%%%%%%%%%%%%%%%%%%%%%%%%%%%%%%%%%%%%%%%%%%%%%%%%%%%%%%%%%%%%%%%%%%%%%%%%%%%%%%%%%%%%%%%%%%%%%%%%%                                        Experiment apparatus
%%%%%%%%%%%%%%%%%%%%%%%%%%%%%%%%%%%%%%%%%%%%%%%%%%%%%%%%%%%%%%%%%%%%%%%%%%%%%%%%%%%%%%%%%%%%%%%%%%%%%%

\section{Experiment apparatus}
\label{sec:expst}
\begin{figure}[h!]
    \centering
    \includegraphics[width = 100 mm]{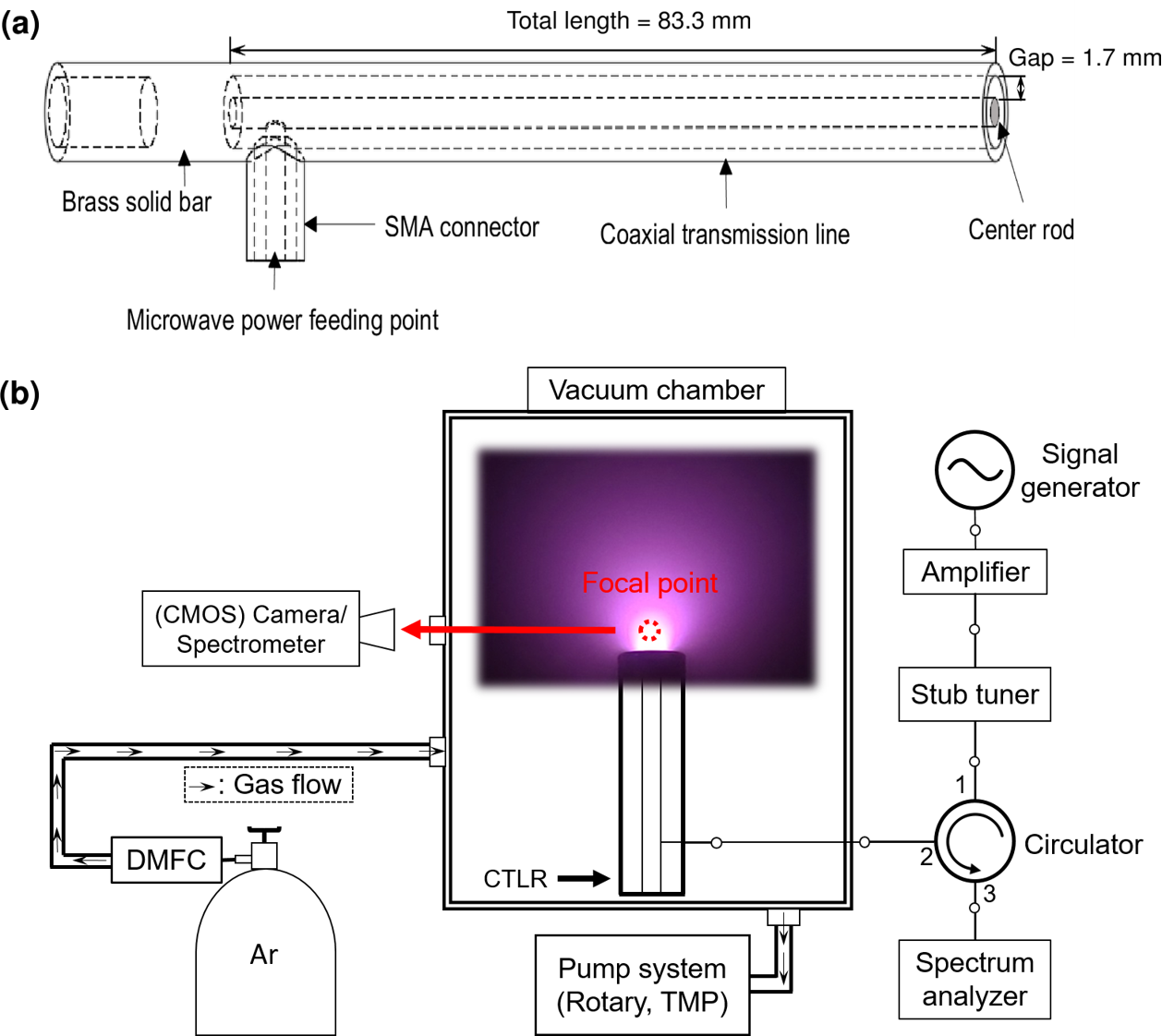}
    \caption{(a): Schematic diagram of the CTLR electrode. The microwave enters the electrode through the side port and propagates to both short- and open-ended sides. The wave reflected at the short end propagates to the open-end and resonates with the original microwave producing a large amplitude. \cite{choi2009microwave}. (b): The schematic diagram of the experimental system. The red dotted circle is the focal point of the collection optics, and its relative coordinate to the electrode is fixed throughout the entire experiment.}
    \label{fig:experimentstructure}
\end{figure}

A coaxial transmission line resonator (CTLR) produces an electrical discharge at the open ended tip of the electrode where the resonance condition is achieved \cite{choi2009microwave}. The designed resonant frequency of the electrode is about 900 MHz, where the wavelength corresponds to one-quarter of the total length of the electrode, $\sim$83.3 mm (Fig.\ref{fig:experimentstructure} (a)), but due to the error and discrepancy in manufacturing, the actual resonance of the electrode appears at 890 MHz based on the S11 measurement by a network analyzer (E8362B, Agilent Technology). Another key parameter of the resonator design is the gap distance between inner- and outer-electrodes. In this research, the gap distance is about 1.7 mm to attain a sufficiently large electric field while matching the characteristic impedance of 50 $\Omega$ in the circuit.

Figure \ref{fig:experimentstructure}(b) shows a schematic of the experimental system. The vacuum chamber was ventilated to 0.2 mTorr and filled with pure argon gas up to the pressure values of the experimental conditions around 20 mTorr. The operating pressure, which is one of the main control parameters for this experiment, was adjusted by a digital mass flow controller (DMFC, MFC KOREA). The net dissipated power by the system is preserved throughout the experiments. The incident power is maintained to 7 W. The reflected microwave is collected via a circulator and is monitored with a spectrum analyzer (E4408B, Agilent Technology) while adjusting a stub tuner to maintain its power level to less than -20 dB of the incident wave. The plasma images were taken with a high resolution CMOS camera. The optical emissions from the plasma were collected by an optical fiber and lens array focusing at the vicinity of the electrode (Fig.\ref{fig:experimentstructure} (b)) and analyzed using a spectrometer (USB2000, Ocean Optics).

\FloatBarrier

%\pagebreak

%%%%%%%%%%%%%%%%%%%%%%%%%%%%%%%%%%%%%%%%%%%%%%%%%%%%%%%%%%%%%%%%%%%%%%%%%%%%%%%%%%%%%%%%%%%%%%%%%%%%%%                                        Experimental results
%%%%%%%%%%%%%%%%%%%%%%%%%%%%%%%%%%%%%%%%%%%%%%%%%%%%%%%%%%%%%%%%%%%%%%%%%%%%%%%%%%%%%%%%%%%%%%%%%%%%%%

\section{Experimental results}
\label{sec:exp}
\subsection{\label{subsec:plasmaplume} Visual characteristics of the plasma discharges}

\begin{figure} [h!]
    \centering
    \includegraphics [width = 164.6 mm] {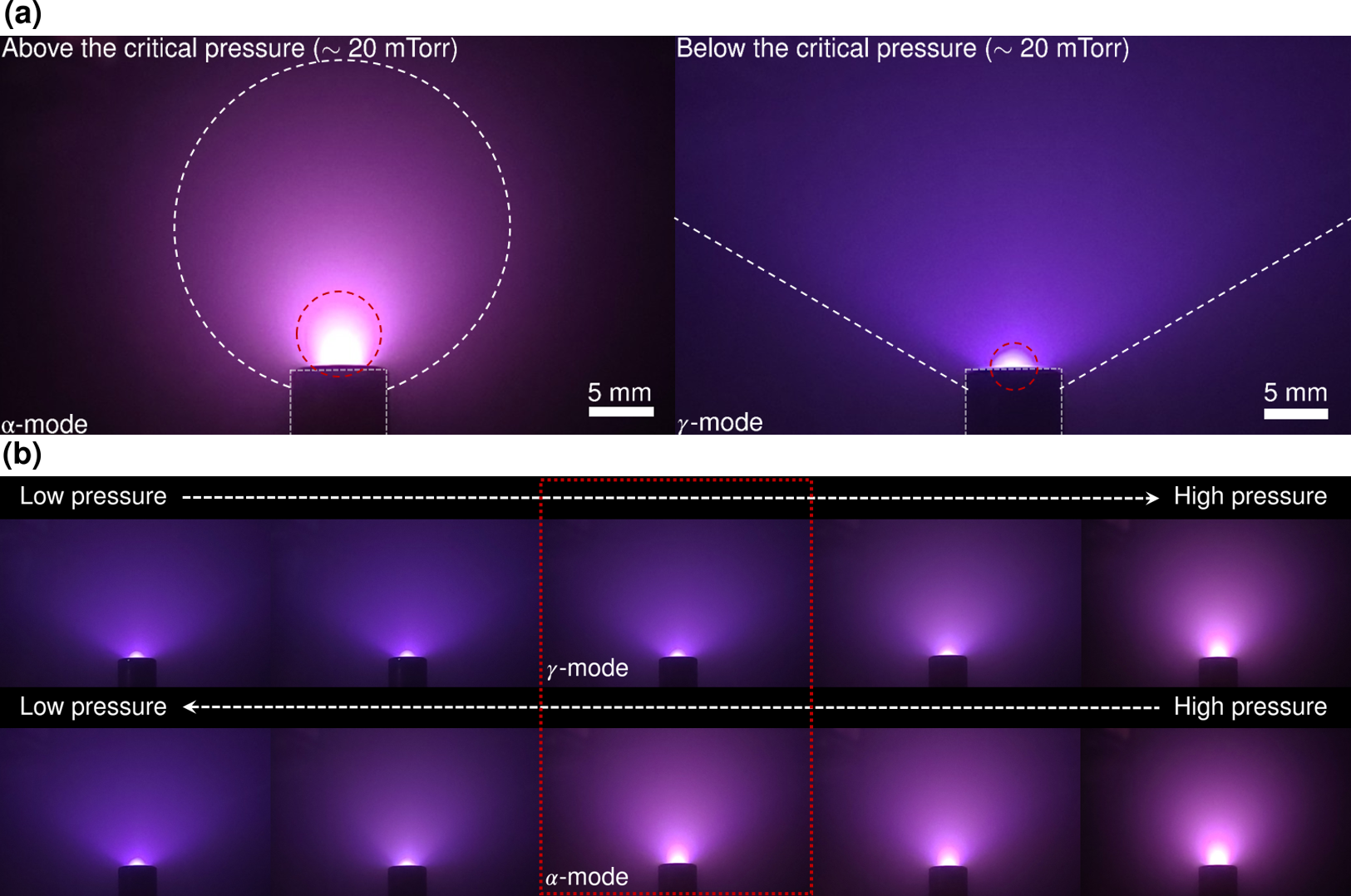}
    \caption{(a): Plasma plumes at two different regimes: ‘alpha mode’ plasma with the gas pressure
above the critical pressure (left) and ‘gamma mode’ plasma with the gas pressure below the critical pressure (right). (b): Change of the plasma plumes with increasing pressure (top) and with decreasing pressure (bottom). These snapshots show the hysteresis in the process of mode transition. The middle images clearly show different states under the same operating pressure.}
    \label{fig:plasmaplumes}
\end{figure}
\FloatBarrier

The visible characteristics of the plasma including volume, shape, and color drastically change when crossing the critical pressure ($P_c$). For a capacitive plasma, it has been reported that when the bulk size reduces and the sheath widens owing to the discharge operating parameters, the plasma becomes dominated by sheaths which can only be sustained in the so-called $\gamma$-mode. On the other hand, when the sheath is narrow, the discharge is dominated by the bulk-plasma electrons that is called $\alpha$-mode \cite{shi2006evolution,liu2009electron}. In this experiment, when $P>P_c$, the bright region of the plasma bulk is large and clear, and the sheath is spherical-shaped, whereas when $P<P_c$, the bulk size reduces, and the diffusive sheath transforms into a reverse cone-shaped (Fig.\ref{fig:plasmaplumes}(a)). Thus, we assign the discharge mode of the plasma at a higher pressure than $P_c$ as the $\alpha$-mode, and that of the plasma at a lower pressure than $P_c$  as the $\gamma$-mode. 

Interestingly, we observe the mode transition occuring at different pressures depending on the direction of the pressure change as shown in Fig.\ref{fig:plasmaplumes}(b). We infer that there exist two different stable states of plasma discharges that correspond to the distinct modes ($\alpha$ and $\gamma$) even for the same operating conditions due to the discrepancy between the critical pressures of the mode transition depending on the pathway of altering pressure. We analyze the optical emissions from the plasma to verify the cause of the hysteresis more descriptively.

\subsection{\label{subsec:electrontemperature} Change in the electron temperature}

\begin{figure}[h!]
    \centering
    \includegraphics[width = 164.6 mm]{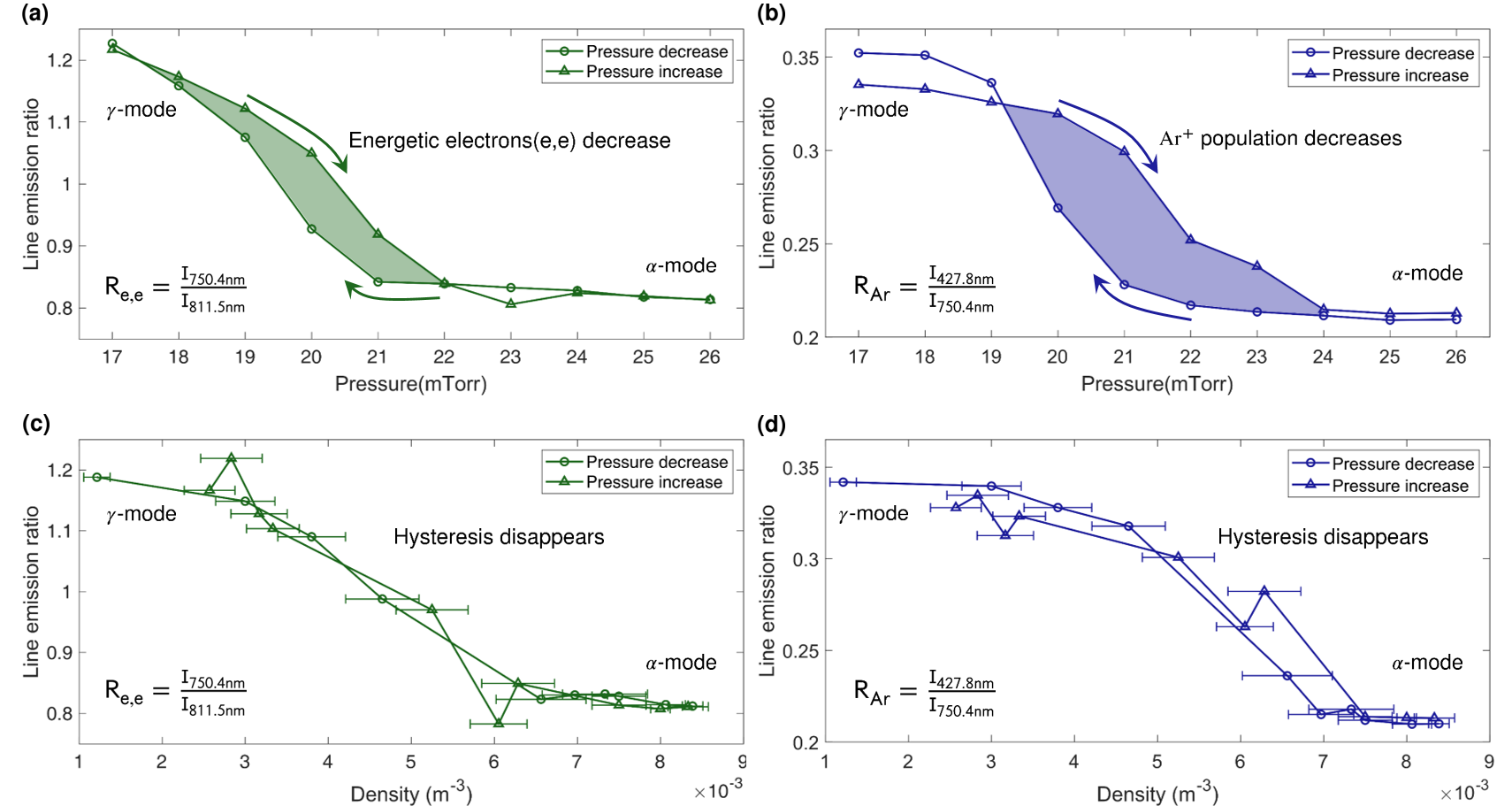}
    \caption{(a): Line emission ratio ($R_{\mathrm{e,e}}$) versus pressure. $R_{\mathrm{e,e}}=\frac{I_{\mathrm{750.4 nm}}}{I_{\mathrm{811.5 nm}}}$  represents the proportion of energetic electrons at the corresponding pressure. The high proportion at pressures below the critical pressure point indicates a high electron temperature. There is a hysteresis depending on the direction of pressure change. (b): Line emission ratio ($R_{\mathrm{Ar}}$) versus pressure. $R_{\mathrm{Ar}}=\frac{I_{\mathrm{427.8 nm}}}{I_{\mathrm{750.4 nm}}}$  expresses the ratio of Ar II emission and Ar I emission at the corresponding pressure. The ratio sharply increases below the critical pressure point and there is a hysteresis depending on the direction of pressure change. (c): The emission line ratio ($R_{\mathrm{e,e}}=\frac{I_{750.4nm}}{I_{811.5nm}} $) versus experimental pressure divided by the measured rotational temperature at that pressure. (d): Line emission ratio ($R_{\mathrm{Ar}}= \frac{I_\mathrm{427.8 nm}}{I_\mathrm{750.4 nm}} $) versus experimental pressure divided by the rotational temperature at that pressure. The hysteresis disappears when expressed as a neutral density in (c) and (d).}
    \label{fig:4plot}
\end{figure}

The discharge mode transition can be identified by examining the change in the electron temperature, because high-energy electrons are essential for the plasma to operate in $\gamma$ mode \cite{lee2017extended,lee2018generation}. The change in the electron temperature can be recognized by the ratio of the emission line with high threshold energy to the emission line with low threshold energy. As the emission line ratio is tracked over a pressure change, the slope indicates an increase in energetic electrons. Specifically, the line emission with a wavelength of 750.4 nm, [$3s^2 3p^5 (^2 P^o_{1/2})4p \rightarrow 3s^23p^5(^2P^o_{1/2})4s$],\cite{NIST_ASD} which is mainly caused by direct excitation of ground-state atoms, was compared with the line emission with a wavelength of 811.5 nm, [$3s^2 3p^5 (^2 P^o_{3/2})4p \rightarrow 3s^2 3p^5 (^2 P^o_{3/2})4s$],\cite{NIST_ASD} which is mainly populated by electron impact excitation from metastable levels. The threshold energies required for excitation are approximately 13.5 eV (750.4 nm) and 2 eV (811.5 nm), respectively \cite{dunnbier2015stability}. We describe the profiles of the emission line intensity ratios ($R_{\mathrm{e,e}}=\frac{I_{\mathrm{750.4 nm}}}{I_{\mathrm{811.5 nm}}}$) versus the pressure depending on the direction of pressure change (increasing and decreasing). (Figure.\ref{fig:4plot}(a))

\FloatBarrier

The measurement result (Fig.\ref{fig:4plot}(a)) illustrates that the ratio of energetic electrons increases rapidly below the critical pressure ($P_\mathrm{c}$) indicating that the population of energetic electrons increases rapidly. The sudden increase in the ratio of energetic electrons in the sustained plasma indicates that the plasma discharge mode transition occurs from the $\alpha$ to the $\gamma$ mode. The results of Fu et al.\cite{fu2020similarity1,fu2020similarity2} validate the discharge mode transition in our experiment. Fu et al.\cite{fu2020similarity1,fu2020similarity2} reported that the similarity law was identified in similar discharge structures through driving frequency scaling. It was shown that the electron energy probability function (EEPF) overlapped when the driving frequency was varied in the $\alpha$ mode, demonstrating that in a similar discharge mode, the proportion of energetic electrons is maintained. However, Fu et al.\cite{fu2020similarity2} verified that a state in which the similarity law is violated occurs when the driving frequency is below the critical frequency. The EEPF has a more pronounced profile in the high-energy region, describing both the increase in the proportion of energetic electrons and high electron temperature. The simulation results of Fu et al.\cite{fu2020similarity1,fu2020similarity2} verified that many high-energy electrons are generated in the other discharge mode (different from the $\alpha$ mode) where the similarity is violated, confirming that the discharge mode transition occurs in our experiment. Lee et al.\cite{lee2017extended,lee2017scalings,lee2018generation} intuitively explained that such a transition is the discharge mode transition from the $\alpha$ mode to $\gamma$ mode in terms of electron confinement (mentioned in the introduction section). The experimental result (Fig.\ref{fig:4plot}(a)) is in good agreement with the previous literature,\cite{lee2017extended,lee2017scalings,lee2018generation,fu2020similarity1,fu2020similarity2} showing the ($\alpha - \gamma$) discharge mode transition.

Most importantly, the increasing proportion of energetic electrons depending on the direction of the pressure change is considerably different, as observed in the previous visual image results. Thus, it clearly demonstrates that hysteresis occurs across the discharge mode transition depending on the direction of the pressure change. Additionally, this study identifies the difference in the population of $\mathrm{Ar^+}$ ions between the $\gamma$- and $\alpha$-modes.

\subsection{\label{subsec:argonion} Change in the population of $\mathrm{Ar^+}$ ions}

In the $\gamma$ mode, we predicted that sufficient ions for the $\gamma$ process were generated by ionization owing to the numerous energetic electrons. In actual OES measurements, it was observed that the distribution of Ar II emission lines in the $\gamma$ mode was more pronounced overall than in the $\alpha$ mode. In order to clearly investigate the difference in intensity of the Ar II emission line between the $\alpha$ and the $\gamma$ modes, two different peaks were selected for comparison; the intensities of the Ar I emission lines and Ar II emission lines. Specifically, the emission line with a wavelength of 427.8 nm, [$3s^2 3p^4 (^1 D)4p \rightarrow 3s^2 3p^4 (^1 D)4s$],\cite{NIST_ASD} which is a representative line among Ar II lines, was normalized to the emission line with a wavelength of 750.4 nm, [$3s^2 3p^5 (^2 P^o_{1/2})4p \rightarrow 3s^2 3p^5 (^2 P^o_{1/2})4s$],\cite{NIST_ASD} which is a representative Ar I line and one of the emission lines with the strongest intensity in the measurement. We describe the profile of the emission line intensity ratio ($R_{\mathrm{Ar}}=\frac{I_{\mathrm{427.8 nm}}}{I_{\mathrm{750.4 nm}}} $) versus pressure depending on the direction of pressure change (increasing and decreasing). (Figure \ref{fig:4plot}(b))

The result (Fig.\ref{fig:4plot}(b)) shows that the proportion of Ar II emission increases rapidly as the pressure falls below the critical pressure point ($P_{\mathrm{c}} \sim 20\:\mathrm{mTorr}$), indicating an increase in the population of argon ions ($\mathrm{Ar^+}$). To produce more ions, it is necessary to generate more energetic electrons with high kinetic energy that can trigger the ionization process inside the plasma. Therefore, the abrupt increase in the proportion of argon ions ($\mathrm{Ar^+}$) is an important result to illustrate that the discharge mode transition occurs from the $\alpha$ mode to the $\gamma$ mode.

In addition, the population profile of argon ions clearly reveals a hysteresis in the pressure of the discharge mode transition depending on the direction of the pressure change, as observed both in the visual image (Fig.\ref{fig:plasmaplumes} (b)) and the population profile of energetic electrons (Fig.\ref{fig:4plot}(a)). The discovery of hysteresis during the mode transition is a remarkably valuable result because it suggests that the plasma can have two different stable states with separate properties depending on the direction of pressure change in the transition region, especially notable for many industries requiring plasma processing (semiconductor manufacturing, etc.). 

\subsection{\label{subsec:eli_hys} The elimination of hysteresis}

We inferred that the primary cause of the hysteresis is the sudden increase in argon ions resulting from the mode transition. Therefore, we hypothesized the phenomenon caused by the numerous argon ions in the $\gamma$ mode. To verify the cause of the hysteresis, we estimated the rotational and vibrational temperatures through optical diagnosis and simulation data from the open-source LIFBASE \cite{luque1999lifbase}. As a result, Figure \ref{fig:4plot}(c) and (d) reveal that the hysteresis generated in the mode transition regime can be eliminated when the optical diagnosis data are reorganized for the neutral density considering the measured rotational temperature introducing the ideal gas law.

%\pagebreak

%%%%%%%%%%%%%%%%%%%%%%%%%%%%%%%%%%%%%%%%%%%%%%%%%%%%%%%%%%%%%%%%%%%%%%%%%%%%%%%%%%%%%%%%%%%%%%%%%%%%%%                                        Disscussion and Analysis
%%%%%%%%%%%%%%%%%%%%%%%%%%%%%%%%%%%%%%%%%%%%%%%%%%%%%%%%%%%%%%%%%%%%%%%%%%%%%%%%%%%%%%%%%%%%%%%%%%%%%%

\section{Discussion and Analysis}
\label{sec:dis}
\subsection*{\label{subsec:hysteresis} The hysteresis of the rotational and vibrational temperature}

The most notable phenomenon in this experiment is the hysteresis depending on the direction of the pressure change when switching the modes ($\alpha - \gamma$) or ($\gamma - \alpha$). In Fig.\ref{fig:4plot}(b), which shows the ratio of the Ar II emission line, and Fig.\ref{fig:4plot}(a), which shows the ratio of energetic electrons, the critical point is at a higher pressure when increasing the pressure than when decreasing the pressure. The fact that the proportion of Ar ions sharply increases in $\gamma$-mode is particularly important in analyzing the primary cause of this hysteresis. In general, neutral gas heating in microwave plasma mostly occurs through electron-neutral elastic collisions, especially in $\alpha$-mode. In this case, the energy transfer rate is notably slow because the electron mass is much smaller than that of the neutral gas. Therefore, the neutral gas heating effect is negligible. However, fast gas heating occurs under conditions where certain other reactions dominate, such as ion-momentum collisions and charge exchange collisions. Because the mass of ions is almost equal to that of neutral gas, fast ions accelerate in the sheath quickly and transfer significant quantities of kinetic energy to neutral gas through the above reactions \cite{popov2011fast,lee2017evolution}.

Consequently, as the population of Ar ions increases in the $\gamma$-mode, the fast gas heating effect dominates, and the gas temperature inside the plasma will increase sharply. An increase in the gas temperature indicates a decrease in the gas density inside the plasma. Based on this phenomenon, we inferred that the abrupt increase in gas temperature in the $\gamma$ mode provokes the gas density of the $\gamma$ mode to become smaller than that of the $\alpha$ mode. As a result, a higher critical pressure is required for the $\gamma$ to $\alpha$ mode transition when the system pressure increases as compared to the case of the $\alpha$ to $\gamma$ mode transition (system pressure decrease). 

\begin{figure}[h!]
    \centering
    \includegraphics[width = 164.6 mm]{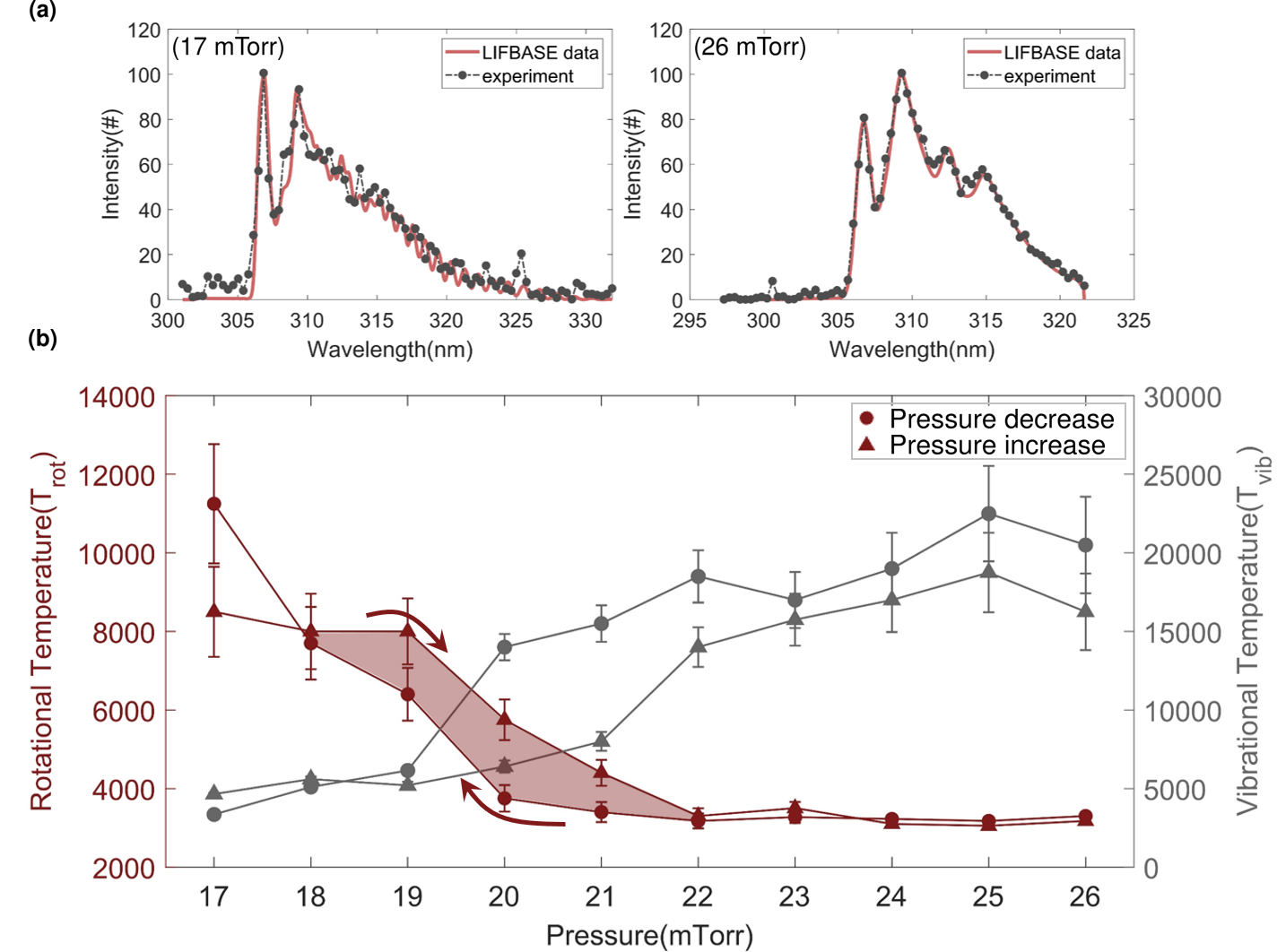}
    \caption{(a): The example fittings between experimental values and LIFBASE data for OH(A-X) emission lines. (b): The rotational temperature and vibrational temperature measured by the OH(A-X) emission lines: the rotational temperature increases sharply below the critical pressure ($P_{\mathrm{c}}$) when reducing the pressure (triangle, red line), and is higher in the direction of increasing pressure than in the direction of decreasing pressure (circle, red line). The vibrational temperature shows a contrary trend (gray lines)}
    \label{fig:temperature}
\end{figure}
\FloatBarrier

This was verified by measuring the rotational temperature of the diatomic molecules inside the plasma. The experiment used pure argon gas but contained impurities on the order of 1\% (base vacuum pressure was approximately 1\% of the total pressure). This impurity originates from the chamber wall and is mostly water vapor. It is possible to measure the OH(A-X) emission line caused by trace amounts of water vapor \cite{laux2003optical}. We estimated the rotational temperature and vibrational temperature by comparing the measured OH(A-X) emission lines with the simulation data provided by the open-source LIFBASE program. To determine the rotational temperature, the OH(A-X) emission lines of the LIFBASE data were fit in good agreement with the experimental results. These branches formed distinct peaks at approximately 307 and 309 nm.\cite{laux2003optical} We also considered the peaks generated from 310 to 320 nm to determine the vibrational temperature. For best fit, the resolution of the LIFBASE data \cite{luque1999lifbase} was set to approximately 0.7 nm (the resolution spec of the experimental measurement equipment is about 0.4 nm). Examples of the fittings are presented in Fig.\ref{fig:temperature}(a).
\\

The experimental results (Fig.\ref{fig:temperature}(b)) reveal that the rotational temperature sharply increases when switching from the $\alpha$ mode to the $\gamma$ mode, and in the range where hysteresis occurs ( from approximately 18 mTorr to 22 mTorr), the rotational temperature remains higher in the direction of increasing pressure than in the direction of decreasing pressure, in agreement with the hypothesis. Likewise, in the case of the vibrational temperature, it is lower in the increasing direction than in the decreasing direction, resulting in the opposite tendency. To verify this hypothesis, we analyzed the relationship between individual temperatures (rotational and vibrational) and physical parameters of plasma utilizing basic models reported by Kang et al \cite{zheng2002molecular} and Lavrov et al \cite{lavrov1978relation}.

\subsubsection{\label{subsection:vibrational hysteresis} Hysteresis of $T_\mathrm{vib}$ between mode transition ($\alpha - \gamma$)}

The vibrational temperature has a hysteresis opposite to the rotational temperature because the vibrational temperature is greatly affected not only by the gas density but also by the electron temperature and electron density \cite{zheng2002molecular}. In general, if the plasma is not in equilibrium, the vibrational and rotational temperatures will not be the same. Because a low-pressure plasma, such as in the experiment, corresponds to a non-equilibrium plasma, it is necessary to know which plasma parameters determine the vibrational temperature. Kang et al \cite{zheng2002molecular} theoretically and experimentally analyzed the result of a rapid rise in the vibrational temperature when the pressure increases in a low-pressure plasma. According to Kang et al \cite{zheng2002molecular}, the following basic assumptions are accepted regarding the experimental conditions: 1. The excited state is generated only from ground-state electron impact excitation. 2. The Franck-Condon principle is applicable to these excitation transitions. 3. The collisional de-excitation process is not significant and introduces a Boltzmann distribution for all vibrational levels of molecules to be measured. In this work, the electron impact is the most dominant collision in the microwave-driven low-temperature plasma, and we observe the transition from the first excited state of OH molecules to the ground-state of OH molecules. The above assumptions are sufficiently applicable for OH (A-X) emission line measurements in the microwave plasma. Therefore, we can describe the concentration($N_{mv}$) of the vibrationally excited OH ($\mathrm{A^{2}}\Sigma^{+}$) molecules with vibrationalnumber \textit{v} in the excited state \textit{m} using Equation (\ref{eq:exvib}).

\begin{equation}
    N_{mv} = N^{B}_{mv} = n_{0}\exp(-\frac{E_{mv}}{kT^{m}_\mathrm{{vib}}})
    \label{eq:exvib}
\end{equation}

Equation (1) shows that this plasma corresponds to the local thermalized system and $T_{\mathrm{vib}}^{m}$ is related to $N_{mv}$ via Boltzmann’s relationship. In addition, we only consider the excited state generated from ground-state electron impact excitation and regard that the observed plasmas are in a steady-state, as expressed by 

\begin{equation}
    \frac{dN_{mv}}{dt} = n_{\mathrm{e}}n_{\mathrm{g}}K_{\mathrm{v}(T_{\mathrm{e}})} - \frac{1}{\Delta t}N_{mv} = 0
    \label{eq:steadystate}
\end{equation}

We can describe the concentration of the vibrationally excited OH ($A^2 \Sigma^+$) molecules ($N_{mv}$) using Equation (\ref{eq:exvib2}).

\begin{equation}
    N_{mv} = n_{\mathrm{e}}n_{\mathrm{g}}K_{\mathrm{v}(T_{\mathrm{e}})}\Delta t
    \label{eq:exvib2}
\end{equation}

where $n_{\mathrm{e}}$ is the electron density, $K_{\mathrm{v}(T_{\mathrm{e}})}$ is the vibrational excitation rate as a function of electron temperature ($T_{\mathrm{e}}$), $n_{\mathrm{g}}$ is the background gas density, and $\Delta t$ is the time until spontaneous de-excitation occurs. Consequently, we can define the relation between $T_{\mathrm{vib}}^{m}$ and several parameters using equations (\ref{eq:exvib}) and (\ref{eq:exvib2}) as follows:

\begin{equation}
    T_{\mathrm{vib}}^{m} = (\frac{E_{mv}}{k})/\ln(\frac{n_0}{n_{\mathrm{e}}n_{\mathrm{g}}K_{\mathrm{v}(T_{\mathrm{e}})\Delta t}})
    \label{eq:vibT}
\end{equation}

\begin{figure}[h!]
    \centering
    \includegraphics[width = 110 mm]{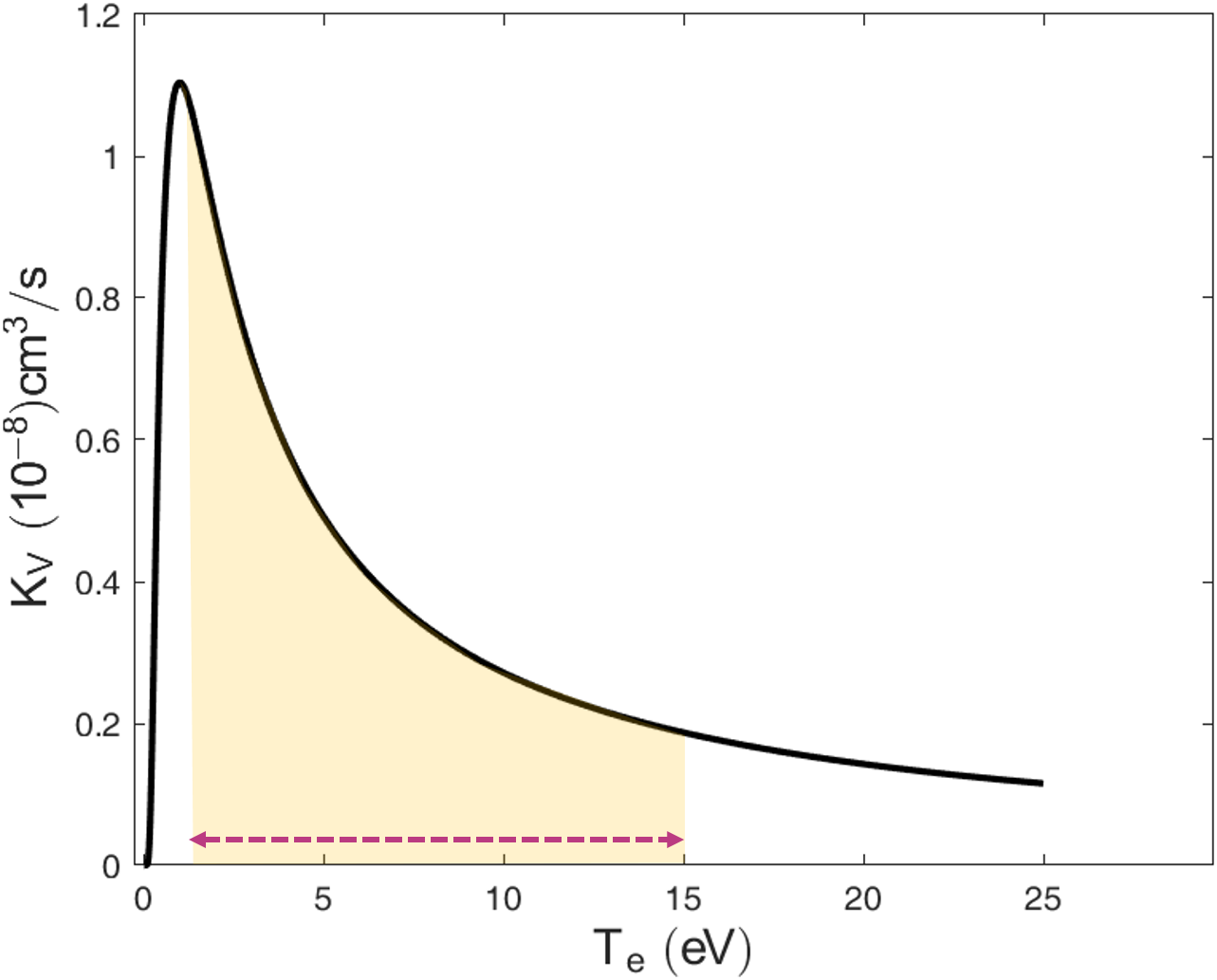}
    \caption{Vibrational excitation rate versus electron temperature \cite{lieberman2005principles}}
    \label{fig:vib_exc}
\end{figure}

As shown in Equation (\ref{eq:vibT}), the vibrational temperature is affected by the vibrational excitation rate and electron density. The vibrational excitation rate is sensitively affected by changes in the electron temperature below approximately 20 eV, as shown in Figure \ref{fig:vib_exc}. 

\FloatBarrier

In our experiment, although a change in both the gas density and the electron temperature was observed during the mode transition process ($\gamma$ - $\alpha$), the change in the electron temperature more dramatically affected the vibrational temperature than the increase in the gas density did. This is because the change in the electron temperature results in a large change in the vibrational excitation rate, leading to a significant variation in the vibrational temperature. According to (\ref{eq:vibT}), the vibrational temperature tends to be inversely proportional to the electron temperature. 

As a result, a sudden change in the electron temperature induces an abrupt change in the vibrational temperature. This analysis demonstrates the discharge mode transition, according to how the vibrational temperature profile exhibits an abrupt change in electron temperature; the hysteresis of the electron temperature causes an inverse hysteresis of the vibrational temperature by the same relationship. The aim of estimating the rotational and vibrational temperatures was to evaluate the translational gas temperature profile. This was achieved by analyzing the rotational temperature.

\subsubsection{\label{rotational hysteresis} Hysteresis of $T_{\mathrm{rot}}$ between mode transition ($\alpha - \gamma$)}

As mentioned earlier, because a low-pressure plasma is a non-equilibrium plasma, the rotational temperature is not directly equal to the gas temperature, but is dependent upon it \cite{bruggeman2014gas,lavrov1978relation}. Lavrov et al \cite{lavrov1978relation} reported the relationship between rotational temperature and gas temperature in low-pressure plasma through theoretical analysis and experimental verification. Let $\nu_{\mathrm{g}}$ be the rate at which the excited molecule collides with the ground-state molecule, and let the probability that the excited molecule decays be $\nu_{\mathrm{r}}$. When $\nu_{\mathrm{g}} > \nu_{\mathrm{r}}$, the rotational temperature ($T_{\mathrm{rot}}$) is approximately equal to the translational temperature ($T_{\mathrm{g}}$). However, low-pressure  plasma corresponds to $\nu_{\mathrm{g}} \leq \nu_{\mathrm{r}}$. In general, when the Boltzmann distribution is satisfied, the population ($N_{mvJ}$) with vibrational quantum number ($v$) and rotational quantum number ($J$) in the excited electronic state ($m$) is described as 

\begin{equation}
    N_{mvJ} = N_{mv0}(2J+1)\exp(-\frac{E_{mvJ}}{kT_{\mathrm{rot}}^{mv}})
    \label{eq:rotex}
\end{equation}

When $\nu_{\mathrm{r}} \geq \nu_{\mathrm{g}},\nu_{\mathrm{e}}$ (the frequency with which excited molecules collide with electrons), $N_{mvJ}$ is governed by the balance between excitation and decay of the molecules in the plasma if the following three assumptions are applicable: (1) The population of rotational sublevels of the electronic and vibrational ground state ($n, 0$), obeys a Boltzmann distribution with temperature ($T_{\mathrm{rot}}^{n0}$). (2) The excited states ($m,v$) are populated by electron impact from the ($n,0$) state and are de-populated by spontaneous decay. (3) No change in the angular momentum of the molecule ($\Delta J=0$) occurs during excitation and the excitation cross sections are independent of $J$. These assumptions are similar to those used in the previous analysis (see Discussion 1.). Therefore, accepting that the above assumptions are applicable to our experiment, the relation is expressed as: 

\begin{equation}
    N_{mvJ} = N_{mv0}(2J+1)\exp(-\frac{E_{n0J}}{kT_{\mathrm{rot}}^{n0}})
    \label{eq:rotex2}
\end{equation}

Comparing equations (\ref{eq:rotex}) and (\ref{eq:rotex2}), assuming $E_{n0J} \sim J(J+1)$, we can deduce the following:

\begin{equation}
    T_{rot}^{mv} = \frac{B_{mv}}{B_{n0}}T_{\mathrm{rot}}^{n0}
    \label{eq:rot}
\end{equation}

where $B_{mv}$ and $B_{n0}$ are rotational constants. The distribution over rotational sublevels of the electronic-vibrational ground state is established under the influence of various processes; collisions of molecules with each other, with electrons, and with the walls of the discharge tube. Thus, this distribution is a Boltzmann distribution of $T_{\mathrm{rot}}^{n0} = T_{\mathrm{g}}$ for the same rotational levels when $\lambda / R \ll 1$ is satisfied, where $\lambda$ is the molecular mean free path and R is the discharge tube radius. Finally, we can determine $T_{\mathrm{g}}$ from the measured rotational temperature of various excited states, as shown in Equation (\ref{eq:gas}).

\setlength{\belowdisplayskip}{0pt} \setlength{\belowdisplayshortskip}{0pt}
\setlength{\abovedisplayskip}{0pt} \setlength{\abovedisplayshortskip}{0pt}
\begin{equation}
    T_{\mathrm{rot}}^{mv} = \frac{B_{mv}}{B_{n0}}T_{\mathrm{g}}
    \label{eq:gas}
\end{equation}

The actual plasma internal gas temperature follows the shape of the rotational temperature profile measured in the experiment. In this study, the abrupt increase in rotational temperature (Fig.\ref{fig:temperature}(b)) demonstrates that the fast gas heating effect occurs strongly when transitioning to the $\gamma$ mode.

As a result, the gas density in the plasma is rapidly lowered, maintaining a lower gas density in the $\gamma$ mode than in the $\alpha$ mode, until pressure meets the mode transition regime (around 20 mTorr) in the direction of increasing pressure. Because of this phenomenon, even if the chamber pressure increases again and reaches the critical pressure observed in the direction of decreasing pressure, the effective gas density does not reach the critical point owing to the gas heating effect. As a result, a higher pressure is required for the mode transition ($\gamma - \alpha$) to occur in the reversed direction.

To verify this analysis more conveniently, we reorganized the data from emission line ratio versus pressure into emission ratio versus density. In our experiment, all variables except the pressure variable were fixed, so if the effective pressure (effective density) was the same, it indicates that the plasma was in the same state. Hence, the plasma in the same state has the same Ar I and Ar II emission line ratios. Accordingly, we converted Fig.\ref{fig:4plot}(a) and Fig.\ref{fig:4plot}(b), which show the emission line ratio versus pressure, to the emission line ratio versus effective density using the measured rotational temperature data, as shown in Fig.\ref{fig:4plot}(c),(d).
 
Figure \ref{fig:4plot} (c) and (d) shows high agreement between two profiles, with the direction of decreasing pressure and the direction of the increasing pressure, respectively. This fact strongly supports the analysis of the hysteresis between the mode transition ($\alpha-\gamma$).

%\pagebreak

%%%%%%%%%%%%%%%%%%%%%%%%%%%%%%%%%%%%%%%%%%%%%%%%%%%%%%%%%%%%%%%%%%%%%%%%%%%%%%%%%%%%%%%%%%%%%%%%%%%%%%                                           Conclusion
%%%%%%%%%%%%%%%%%%%%%%%%%%%%%%%%%%%%%%%%%%%%%%%%%%%%%%%%%%%%%%%%%%%%%%%%%%%%%%%%%%%%%%%%%%%%%%%%%%%%%%

\section{Conclusion}
\label{sec:con}
We found hysteresis in microwave-driven low-temperature plasma across the discharge mode transition ($\alpha - \gamma$). For the same global operation parameters (pressure, power, and gas composition), the discharge mode differs depending on the direction of the pressure change. This observation suggests that plasma under the same operating parameters can have two different stable states. The experiment was conducted for argon plasma generated by a cylindrical resonator electrode at 900 MHz with varying pressure. We have identified the pressure at which the discharge mode change occurs by observing two separate plasma properties related to the electron kinetics that depend on the neutral particle density and the electron temperature: the plasma volume and the ratio of two emission peaks from different excitation processes. We verified that the discharge mechanism is different by comparing the peak emitted owing to energetic electrons with the peak emitted owing to low-energy electrons. Significant changes in the vibrational temperature also strongly supported this analysis.

The hysteresis across the discharge mode transition was interpreted as the difference in the gas density of the plasma, caused by the fast gas heating effect in the $\gamma$ mode. Owing to the fast gas heating in the $\gamma$ mode, the neutral density of the $\gamma$ mode stays smaller than that of the $\alpha$ mode, and the amplitude of the collective electron oscillation in the $\gamma$ mode becomes larger than that of the $\alpha$ mode. As a result, a higher critical pressure is required for the $\gamma$ to $\alpha$ transition when the system pressure increases, compared to the case of the $\alpha$ to $\gamma$ transition when the system pressure decreases. The results of measuring the rotational temperature of diatomic molecules inside the plasma validated this analysis. Thus, for the same global operation parameters, the difference in the gas density of the plasma allows the plasma to have two different steady states under the same global operating conditions.

Modern high-precision semiconductor processes require a uniform and explicit plasma state with consistent properties for accurate semiconductor fabrication. In one semiconductor processing recipe, plasma with varying pressure may be applied to the recipe several times, because the required operating conditions are different for various plasma processes. Therefore, we suggest that the results of this research have significant value in the field of high-precision semiconductor processing because the duplicity of plasma states with different properties (especially ion energy and ion population) under equal global operating parameters (pressure, gas composition, and power) can cause defects in a single cycle of process (a ‘recipe’) with varying directions of pressure. Furthermore, we propose a method of avoiding the hysteresis to ensure the reproducibility of the plasma characteristics, that is, to monitor the neutral density by examining the rotational temperature simultaneously with the operating pressure.

%\setlength{\parskip}{0.1 mm}

%\pagebreak

%%%%%%%%%%%%%%%%%%%%%%%%%%%%%%%%%%%%%%%%%%%%%%%%%%%%%%%%%%%%%%%%%%%%%%%%%%%%%%%%%%%%%%%%%%%%%%%%%%%%%%                                  Supplementary data and acknowledgment
%%%%%%%%%%%%%%%%%%%%%%%%%%%%%%%%%%%%%%%%%%%%%%%%%%%%%%%%%%%%%%%%%%%%%%%%%%%%%%%%%%%%%%%%%%%%%%%%%%%%%%

\section{Acknowledgment}
\label{sec:ack}
This work was supported by Samsung Electronics Co., Ltd (IO201209-07922-01) and the BK21 FOUR Program for Education, Program for Advanced Nuclear Convergence for Future Society of the National Research Foundation of Korea (NRF) grant funded by the Korean government (MSIP).

%\pagebreak

%\section{Supplementary data}
%\label{sec:sup}
%\input{supplementary}

%%%%%%%%%%%%%%%%%%%%%%%%%%%%%%%%%%%%%%%%%%%%%%%%%%%%%%%%%%%%%%%%%%%%%%%%%%%%%%%%%%%%%%%%%%%%%%%%%%%%%%                                              Reference
%%%%%%%%%%%%%%%%%%%%%%%%%%%%%%%%%%%%%%%%%%%%%%%%%%%%%%%%%%%%%%%%%%%%%%%%%%%%%%%%%%%%%%%%%%%%%%%%%%%%%%

%\nocite{*}
%\bibliography{reference}% Produces the bibliography via BibTeX.

\begin{thebibliography}{31}%
\makeatletter
\providecommand \@ifxundefined [1]{%
 \@ifx{#1\undefined}
}%
\providecommand \@ifnum [1]{%
 \ifnum #1\expandafter \@firstoftwo
 \else \expandafter \@secondoftwo
 \fi
}%
\providecommand \@ifx [1]{%
 \ifx #1\expandafter \@firstoftwo
 \else \expandafter \@secondoftwo
 \fi
}%
\providecommand \natexlab [1]{#1}%
\providecommand \enquote  [1]{``#1''}%
\providecommand \bibnamefont  [1]{#1}%
\providecommand \bibfnamefont [1]{#1}%
\providecommand \citenamefont [1]{#1}%
\providecommand \href@noop [0]{\@secondoftwo}%
\providecommand \href [0]{\begingroup \@sanitize@url \@href}%
\providecommand \@href[1]{\@@startlink{#1}\@@href}%
\providecommand \@@href[1]{\endgroup#1\@@endlink}%
\providecommand \@sanitize@url [0]{\catcode `\\12\catcode `\$12\catcode
  `\&12\catcode `\#12\catcode `\^12\catcode `\_12\catcode `\%12\relax}%
\providecommand \@@startlink[1]{}%
\providecommand \@@endlink[0]{}%
\providecommand \url  [0]{\begingroup\@sanitize@url \@url }%
\providecommand \@url [1]{\endgroup\@href {#1}{\urlprefix }}%
\providecommand \urlprefix  [0]{URL }%
\providecommand \Eprint [0]{\href }%
\providecommand \doibase [0]{https://doi.org/}%
\providecommand \selectlanguage [0]{\@gobble}%
\providecommand \bibinfo  [0]{\@secondoftwo}%
\providecommand \bibfield  [0]{\@secondoftwo}%
\providecommand \translation [1]{[#1]}%
\providecommand \BibitemOpen [0]{}%
\providecommand \bibitemStop [0]{}%
\providecommand \bibitemNoStop [0]{.\EOS\space}%
\providecommand \EOS [0]{\spacefactor3000\relax}%
\providecommand \BibitemShut  [1]{\csname bibitem#1\endcsname}%
\let\auto@bib@innerbib\@empty
%</preamble>
\bibitem [{\citenamefont {Moisan}\ \emph {et~al.}(1991)\citenamefont {Moisan},
  \citenamefont {Barbeau}, \citenamefont {Claude}, \citenamefont {Ferreira},
  \citenamefont {Margot}, \citenamefont {Paraszczak}, \citenamefont {Sá},
  \citenamefont {Sauvé},\ and\ \citenamefont {Wertheimer}}]{MM}%
  \BibitemOpen
  \bibfield  {author} {\bibinfo {author} {\bibfnamefont {M.}~\bibnamefont
  {Moisan}}, \bibinfo {author} {\bibfnamefont {C.}~\bibnamefont {Barbeau}},
  \bibinfo {author} {\bibfnamefont {R.}~\bibnamefont {Claude}}, \bibinfo
  {author} {\bibfnamefont {C.~M.}\ \bibnamefont {Ferreira}}, \bibinfo {author}
  {\bibfnamefont {J.}~\bibnamefont {Margot}}, \bibinfo {author} {\bibfnamefont
  {J.}~\bibnamefont {Paraszczak}}, \bibinfo {author} {\bibfnamefont {A.~B.}\
  \bibnamefont {Sá}}, \bibinfo {author} {\bibfnamefont {G.}~\bibnamefont
  {Sauvé}},\ and\ \bibinfo {author} {\bibfnamefont {M.~R.}\ \bibnamefont
  {Wertheimer}},\ }\bibfield  {title} {\bibinfo {title} {Radio frequency or
  microwave plasma reactors? factors determining the optimum frequency of
  operation},\ }\href {https://doi.org/10.1116/1.585795} {\bibfield  {journal}
  {\bibinfo  {journal} {Journal of Vacuum Science \& Technology B:
  Microelectronics and Nanometer Structures Processing, Measurement, and
  Phenomena}\ }\textbf {\bibinfo {volume} {9}},\ \bibinfo {pages} {8} (\bibinfo
  {year} {1991})},\ \Eprint
  {https://arxiv.org/abs/https://avs.scitation.org/doi/pdf/10.1116/1.585795}
  {https://avs.scitation.org/doi/pdf/10.1116/1.585795} \BibitemShut {NoStop}%
\bibitem [{\citenamefont {Musil}(1986)}]{musil1986microwave}%
  \BibitemOpen
  \bibfield  {author} {\bibinfo {author} {\bibfnamefont {J.}~\bibnamefont
  {Musil}},\ }\bibfield  {title} {\bibinfo {title} {Microwave plasma: its
  characteristics and applications in thin film technology},\ }\href@noop {}
  {\bibfield  {journal} {\bibinfo  {journal} {Vacuum}\ }\textbf {\bibinfo
  {volume} {36}},\ \bibinfo {pages} {161} (\bibinfo {year} {1986})}\BibitemShut
  {NoStop}%
\bibitem [{\citenamefont {Vesel}\ and\ \citenamefont
  {Mozetic}(2017)}]{vesel2017new}%
  \BibitemOpen
  \bibfield  {author} {\bibinfo {author} {\bibfnamefont {A.}~\bibnamefont
  {Vesel}}\ and\ \bibinfo {author} {\bibfnamefont {M.}~\bibnamefont
  {Mozetic}},\ }\bibfield  {title} {\bibinfo {title} {New developments in
  surface functionalization of polymers using controlled plasma treatments},\
  }\href@noop {} {\bibfield  {journal} {\bibinfo  {journal} {Journal of Physics
  D: Applied Physics}\ }\textbf {\bibinfo {volume} {50}},\ \bibinfo {pages}
  {293001} (\bibinfo {year} {2017})}\BibitemShut {NoStop}%
\bibitem [{\citenamefont {F{\"o}rch}\ \emph {et~al.}(2005)\citenamefont
  {F{\"o}rch}, \citenamefont {Zhang},\ and\ \citenamefont
  {Knoll}}]{forch2005soft}%
  \BibitemOpen
  \bibfield  {author} {\bibinfo {author} {\bibfnamefont {R.}~\bibnamefont
  {F{\"o}rch}}, \bibinfo {author} {\bibfnamefont {Z.}~\bibnamefont {Zhang}},\
  and\ \bibinfo {author} {\bibfnamefont {W.}~\bibnamefont {Knoll}},\ }\bibfield
   {title} {\bibinfo {title} {Soft plasma treated surfaces: tailoring of
  structure and properties for biomaterial applications},\ }\href@noop {}
  {\bibfield  {journal} {\bibinfo  {journal} {Plasma processes and polymers}\
  }\textbf {\bibinfo {volume} {2}},\ \bibinfo {pages} {351} (\bibinfo {year}
  {2005})}\BibitemShut {NoStop}%
\bibitem [{\citenamefont {Lebedev}(2015)}]{lebedev2015microwave}%
  \BibitemOpen
  \bibfield  {author} {\bibinfo {author} {\bibfnamefont {Y.~A.}\ \bibnamefont
  {Lebedev}},\ }\bibfield  {title} {\bibinfo {title} {Microwave discharges at
  low pressures and peculiarities of the processes in strongly non-uniform
  plasma},\ }\href@noop {} {\bibfield  {journal} {\bibinfo  {journal} {Plasma
  Sources Science and Technology}\ }\textbf {\bibinfo {volume} {24}},\ \bibinfo
  {pages} {053001} (\bibinfo {year} {2015})}\BibitemShut {NoStop}%
\bibitem [{\citenamefont {Donnelly}\ and\ \citenamefont
  {Kornblit}(2013)}]{donnelly2013plasma}%
  \BibitemOpen
  \bibfield  {author} {\bibinfo {author} {\bibfnamefont {V.~M.}\ \bibnamefont
  {Donnelly}}\ and\ \bibinfo {author} {\bibfnamefont {A.}~\bibnamefont
  {Kornblit}},\ }\bibfield  {title} {\bibinfo {title} {Plasma etching:
  Yesterday, today, and tomorrow},\ }\href@noop {} {\bibfield  {journal}
  {\bibinfo  {journal} {Journal of Vacuum Science \& Technology A: Vacuum,
  Surfaces, and Films}\ }\textbf {\bibinfo {volume} {31}},\ \bibinfo {pages}
  {050825} (\bibinfo {year} {2013})}\BibitemShut {NoStop}%
\bibitem [{\citenamefont {Bera}\ \emph {et~al.}(2008)\citenamefont {Bera},
  \citenamefont {Mahata},\ and\ \citenamefont {Maiti}}]{bera2008reliability1}%
  \BibitemOpen
  \bibfield  {author} {\bibinfo {author} {\bibfnamefont {M.}~\bibnamefont
  {Bera}}, \bibinfo {author} {\bibfnamefont {C.}~\bibnamefont {Mahata}},\ and\
  \bibinfo {author} {\bibfnamefont {C.}~\bibnamefont {Maiti}},\ }\bibfield
  {title} {\bibinfo {title} {Reliability of ultra-thin titanium dioxide (tio2)
  films on strained-si},\ }\href@noop {} {\bibfield  {journal} {\bibinfo
  {journal} {Thin Solid Films}\ }\textbf {\bibinfo {volume} {517}},\ \bibinfo
  {pages} {27} (\bibinfo {year} {2008})}\BibitemShut {NoStop}%
\bibitem [{\citenamefont {Bera}\ and\ \citenamefont
  {Maiti}(2008)}]{bera2008reliability2}%
  \BibitemOpen
  \bibfield  {author} {\bibinfo {author} {\bibfnamefont {M.}~\bibnamefont
  {Bera}}\ and\ \bibinfo {author} {\bibfnamefont {C.}~\bibnamefont {Maiti}},\
  }\bibfield  {title} {\bibinfo {title} {Reliability of ultra thin zro2 films
  on strained-si},\ }\href@noop {} {\bibfield  {journal} {\bibinfo  {journal}
  {Microelectronics Reliability}\ }\textbf {\bibinfo {volume} {48}},\ \bibinfo
  {pages} {682} (\bibinfo {year} {2008})}\BibitemShut {NoStop}%
\bibitem [{\citenamefont {Şengül}\ \emph {et~al.}(2008)\citenamefont
  {Şengül}, \citenamefont {Theis},\ and\ \citenamefont {Ghosh}}]{repeatedly}%
  \BibitemOpen
  \bibfield  {author} {\bibinfo {author} {\bibfnamefont {H.}~\bibnamefont
  {Şengül}}, \bibinfo {author} {\bibfnamefont {T.~L.}\ \bibnamefont
  {Theis}},\ and\ \bibinfo {author} {\bibfnamefont {S.}~\bibnamefont {Ghosh}},\
  }\bibfield  {title} {\bibinfo {title} {Toward sustainable nanoproducts},\
  }\href {https://doi.org/https://doi.org/10.1111/j.1530-9290.2008.00046.x}
  {\bibfield  {journal} {\bibinfo  {journal} {Journal of Industrial Ecology}\
  }\textbf {\bibinfo {volume} {12}},\ \bibinfo {pages} {329} (\bibinfo {year}
  {2008})},\ \Eprint
  {https://arxiv.org/abs/https://onlinelibrary.wiley.com/doi/pdf/10.1111/j.1530-9290.2008.00046.x}
  {https://onlinelibrary.wiley.com/doi/pdf/10.1111/j.1530-9290.2008.00046.x}
  \BibitemShut {NoStop}%
\bibitem [{\citenamefont {Graves}(1994)}]{processingsequence}%
  \BibitemOpen
  \bibfield  {author} {\bibinfo {author} {\bibfnamefont {D.~B.}\ \bibnamefont
  {Graves}},\ }\bibfield  {title} {\bibinfo {title} {Plasma processing},\
  }\href@noop {} {\bibfield  {journal} {\bibinfo  {journal} {IEEE transactions
  on Plasma Science}\ }\textbf {\bibinfo {volume} {22}},\ \bibinfo {pages} {31}
  (\bibinfo {year} {1994})}\BibitemShut {NoStop}%
\bibitem [{\citenamefont {Fu}\ \emph {et~al.}(2020{\natexlab{a}})\citenamefont
  {Fu}, \citenamefont {Zheng}, \citenamefont {Zhang}, \citenamefont {Fan},
  \citenamefont {Verboncoeur},\ and\ \citenamefont {Wang}}]{fu2020similarity1}%
  \BibitemOpen
  \bibfield  {author} {\bibinfo {author} {\bibfnamefont {Y.}~\bibnamefont
  {Fu}}, \bibinfo {author} {\bibfnamefont {B.}~\bibnamefont {Zheng}}, \bibinfo
  {author} {\bibfnamefont {P.}~\bibnamefont {Zhang}}, \bibinfo {author}
  {\bibfnamefont {Q.~H.}\ \bibnamefont {Fan}}, \bibinfo {author} {\bibfnamefont
  {J.~P.}\ \bibnamefont {Verboncoeur}},\ and\ \bibinfo {author} {\bibfnamefont
  {X.}~\bibnamefont {Wang}},\ }\bibfield  {title} {\bibinfo {title} {Similarity
  of capacitive radio-frequency discharges in nonlocal regimes},\ }\href@noop
  {} {\bibfield  {journal} {\bibinfo  {journal} {Physics of Plasmas}\ }\textbf
  {\bibinfo {volume} {27}},\ \bibinfo {pages} {113501} (\bibinfo {year}
  {2020}{\natexlab{a}})}\BibitemShut {NoStop}%
\bibitem [{\citenamefont {Godyak}\ \emph {et~al.}(1992)\citenamefont {Godyak},
  \citenamefont {Piejak},\ and\ \citenamefont
  {Alexandrovich}}]{godyak1992evolution}%
  \BibitemOpen
  \bibfield  {author} {\bibinfo {author} {\bibfnamefont {V.}~\bibnamefont
  {Godyak}}, \bibinfo {author} {\bibfnamefont {R.}~\bibnamefont {Piejak}},\
  and\ \bibinfo {author} {\bibfnamefont {B.}~\bibnamefont {Alexandrovich}},\
  }\bibfield  {title} {\bibinfo {title} {Evolution of the
  electron-energy-distribution function during rf discharge transition to the
  high-voltage mode},\ }\href@noop {} {\bibfield  {journal} {\bibinfo
  {journal} {Physical review letters}\ }\textbf {\bibinfo {volume} {68}},\
  \bibinfo {pages} {40} (\bibinfo {year} {1992})}\BibitemShut {NoStop}%
\bibitem [{\citenamefont {Lee}\ \emph {et~al.}(2017{\natexlab{a}})\citenamefont
  {Lee}, \citenamefont {Lee}, \citenamefont {Lee},\ and\ \citenamefont
  {Yun}}]{lee2017extended}%
  \BibitemOpen
  \bibfield  {author} {\bibinfo {author} {\bibfnamefont {M.~U.}\ \bibnamefont
  {Lee}}, \bibinfo {author} {\bibfnamefont {J.}~\bibnamefont {Lee}}, \bibinfo
  {author} {\bibfnamefont {J.~K.}\ \bibnamefont {Lee}},\ and\ \bibinfo {author}
  {\bibfnamefont {G.~S.}\ \bibnamefont {Yun}},\ }\bibfield  {title} {\bibinfo
  {title} {Extended scaling and paschen law for micro-sized radiofrequency
  plasma breakdown},\ }\href@noop {} {\bibfield  {journal} {\bibinfo  {journal}
  {Plasma Sources Science and Technology}\ }\textbf {\bibinfo {volume} {26}},\
  \bibinfo {pages} {034003} (\bibinfo {year} {2017}{\natexlab{a}})}\BibitemShut
  {NoStop}%
\bibitem [{\citenamefont {Lee}\ \emph {et~al.}(2017{\natexlab{b}})\citenamefont
  {Lee}, \citenamefont {Lee}, \citenamefont {Yun},\ and\ \citenamefont
  {Lee}}]{lee2017scalings}%
  \BibitemOpen
  \bibfield  {author} {\bibinfo {author} {\bibfnamefont {M.~U.}\ \bibnamefont
  {Lee}}, \bibinfo {author} {\bibfnamefont {J.}~\bibnamefont {Lee}}, \bibinfo
  {author} {\bibfnamefont {G.~S.}\ \bibnamefont {Yun}},\ and\ \bibinfo {author}
  {\bibfnamefont {J.~K.}\ \bibnamefont {Lee}},\ }\bibfield  {title} {\bibinfo
  {title} {Scalings and universality for high-frequency excited high-pressure
  argon microplasma},\ }\href@noop {} {\bibfield  {journal} {\bibinfo
  {journal} {The European Physical Journal D}\ }\textbf {\bibinfo {volume}
  {71}},\ \bibinfo {pages} {1} (\bibinfo {year}
  {2017}{\natexlab{b}})}\BibitemShut {NoStop}%
\bibitem [{\citenamefont {Shi}\ and\ \citenamefont
  {Kong}(2006)}]{shi2006evolution}%
  \BibitemOpen
  \bibfield  {author} {\bibinfo {author} {\bibfnamefont {J.}~\bibnamefont
  {Shi}}\ and\ \bibinfo {author} {\bibfnamefont {M.}~\bibnamefont {Kong}},\
  }\bibfield  {title} {\bibinfo {title} {Evolution of discharge structure in
  capacitive radio-frequency atmospheric microplasmas},\ }\href@noop {}
  {\bibfield  {journal} {\bibinfo  {journal} {Physical review letters}\
  }\textbf {\bibinfo {volume} {96}},\ \bibinfo {pages} {105009} (\bibinfo
  {year} {2006})}\BibitemShut {NoStop}%
\bibitem [{\citenamefont {Lee}\ \emph {et~al.}(2018)\citenamefont {Lee},
  \citenamefont {Lee},\ and\ \citenamefont {Yun}}]{lee2018generation}%
  \BibitemOpen
  \bibfield  {author} {\bibinfo {author} {\bibfnamefont {M.~U.}\ \bibnamefont
  {Lee}}, \bibinfo {author} {\bibfnamefont {J.~K.}\ \bibnamefont {Lee}},\ and\
  \bibinfo {author} {\bibfnamefont {G.~S.}\ \bibnamefont {Yun}},\ }\bibfield
  {title} {\bibinfo {title} {Generation of energetic electrons in pulsed
  microwave plasmas},\ }\href@noop {} {\bibfield  {journal} {\bibinfo
  {journal} {Plasma Processes and Polymers}\ }\textbf {\bibinfo {volume}
  {15}},\ \bibinfo {pages} {1700124} (\bibinfo {year} {2018})}\BibitemShut
  {NoStop}%
\bibitem [{\citenamefont {Lavrov}\ and\ \citenamefont
  {Otorbaev}(1978)}]{lavrov1978relation}%
  \BibitemOpen
  \bibfield  {author} {\bibinfo {author} {\bibfnamefont {B.}~\bibnamefont
  {Lavrov}}\ and\ \bibinfo {author} {\bibfnamefont {D.}~\bibnamefont
  {Otorbaev}},\ }\bibfield  {title} {\bibinfo {title} {Relation between the
  rotational temperature and gas temperature in a low-pressure molecular
  plasma},\ }\href@noop {} {\bibfield  {journal} {\bibinfo  {journal}
  {Technical Physics Letters}\ }\textbf {\bibinfo {volume} {4}},\ \bibinfo
  {pages} {574} (\bibinfo {year} {1978})}\BibitemShut {NoStop}%
\bibitem [{\citenamefont {Lee}\ \emph {et~al.}(2013)\citenamefont {Lee},
  \citenamefont {Kim},\ and\ \citenamefont {Chung}}]{LeeHC1}%
  \BibitemOpen
  \bibfield  {author} {\bibinfo {author} {\bibfnamefont {H.-C.}\ \bibnamefont
  {Lee}}, \bibinfo {author} {\bibfnamefont {D.-H.}\ \bibnamefont {Kim}},\ and\
  \bibinfo {author} {\bibfnamefont {C.-W.}\ \bibnamefont {Chung}},\ }\bibfield
  {title} {\bibinfo {title} {Discharge mode transition and hysteresis in
  inductively coupled plasma},\ }\href {https://doi.org/10.1063/1.4809925}
  {\bibfield  {journal} {\bibinfo  {journal} {Applied Physics Letters}\
  }\textbf {\bibinfo {volume} {102}},\ \bibinfo {pages} {234104} (\bibinfo
  {year} {2013})},\ \Eprint
  {https://arxiv.org/abs/https://doi.org/10.1063/1.4809925}
  {https://doi.org/10.1063/1.4809925} \BibitemShut {NoStop}%
\bibitem [{\citenamefont {Lee}\ and\ \citenamefont {Chung}(2015)}]{LeeHC2}%
  \BibitemOpen
  \bibfield  {author} {\bibinfo {author} {\bibfnamefont {H.-C.}\ \bibnamefont
  {Lee}}\ and\ \bibinfo {author} {\bibfnamefont {C.-W.}\ \bibnamefont
  {Chung}},\ }\bibfield  {title} {\bibinfo {title} {Effect of electron energy
  distribution on the hysteresis of plasma discharge: Theory, experiment and
  modeling},\ }\href {https://doi.org/10.1038/srep15254} {\bibfield  {journal}
  {\bibinfo  {journal} {Scientific Reports}\ }\textbf {\bibinfo {volume} {5}},\
  \bibinfo {pages} {15254} (\bibinfo {year} {2015})}\BibitemShut {NoStop}%
\bibitem [{\citenamefont {Choi}\ \emph {et~al.}(2009)\citenamefont {Choi},
  \citenamefont {Iza}, \citenamefont {Do}, \citenamefont {Lee},\ and\
  \citenamefont {Cho}}]{choi2009microwave}%
  \BibitemOpen
  \bibfield  {author} {\bibinfo {author} {\bibfnamefont {J.}~\bibnamefont
  {Choi}}, \bibinfo {author} {\bibfnamefont {F.}~\bibnamefont {Iza}}, \bibinfo
  {author} {\bibfnamefont {H.}~\bibnamefont {Do}}, \bibinfo {author}
  {\bibfnamefont {J.}~\bibnamefont {Lee}},\ and\ \bibinfo {author}
  {\bibfnamefont {M.}~\bibnamefont {Cho}},\ }\bibfield  {title} {\bibinfo
  {title} {Microwave-excited atmospheric-pressure microplasmas based on a
  coaxial transmission line resonator},\ }\href@noop {} {\bibfield  {journal}
  {\bibinfo  {journal} {Plasma Sources Science and Technology}\ }\textbf
  {\bibinfo {volume} {18}},\ \bibinfo {pages} {025029} (\bibinfo {year}
  {2009})}\BibitemShut {NoStop}%
\bibitem [{\citenamefont {Liu}\ \emph {et~al.}(2009)\citenamefont {Liu},
  \citenamefont {Iza},\ and\ \citenamefont {Kong}}]{liu2009electron}%
  \BibitemOpen
  \bibfield  {author} {\bibinfo {author} {\bibfnamefont {D.}~\bibnamefont
  {Liu}}, \bibinfo {author} {\bibfnamefont {F.}~\bibnamefont {Iza}},\ and\
  \bibinfo {author} {\bibfnamefont {M.~G.}\ \bibnamefont {Kong}},\ }\bibfield
  {title} {\bibinfo {title} {Electron avalanches and diffused $\gamma$-mode in
  radio-frequency capacitively coupled atmospheric-pressure microplasmas},\
  }\href@noop {} {\bibfield  {journal} {\bibinfo  {journal} {Applied Physics
  Letters}\ }\textbf {\bibinfo {volume} {95}},\ \bibinfo {pages} {031501}
  (\bibinfo {year} {2009})}\BibitemShut {NoStop}%
\bibitem [{\citenamefont {Kramida}\ \emph {et~al.}(2021)\citenamefont
  {Kramida}, \citenamefont {{Yu.~Ralchenko}}, \citenamefont {Reader},\ and\
  \citenamefont {{and NIST ASD Team}}}]{NIST_ASD}%
  \BibitemOpen
  \bibfield  {author} {\bibinfo {author} {\bibfnamefont {A.}~\bibnamefont
  {Kramida}}, \bibinfo {author} {\bibnamefont {{Yu.~Ralchenko}}}, \bibinfo
  {author} {\bibfnamefont {J.}~\bibnamefont {Reader}},\ and\ \bibinfo {author}
  {\bibnamefont {{and NIST ASD Team}}},\ }\href@noop {} {}\bibinfo
  {howpublished} {{NIST Atomic Spectra Database (ver. 5.9), [Online].
  Available: {\tt{https://physics.nist.gov/asd}} [2017, April 9]. National
  Institute of Standards and Technology, Gaithersburg, MD.}} (\bibinfo {year}
  {2021})\BibitemShut {NoStop}%
\bibitem [{\citenamefont {D{\"u}nnbier}\ \emph {et~al.}(2015)\citenamefont
  {D{\"u}nnbier}, \citenamefont {Becker}, \citenamefont {Iseni}, \citenamefont
  {Bansemer}, \citenamefont {Loffhagen}, \citenamefont {Reuter},\ and\
  \citenamefont {Weltmann}}]{dunnbier2015stability}%
  \BibitemOpen
  \bibfield  {author} {\bibinfo {author} {\bibfnamefont {M.}~\bibnamefont
  {D{\"u}nnbier}}, \bibinfo {author} {\bibfnamefont {M.}~\bibnamefont
  {Becker}}, \bibinfo {author} {\bibfnamefont {S.}~\bibnamefont {Iseni}},
  \bibinfo {author} {\bibfnamefont {R.}~\bibnamefont {Bansemer}}, \bibinfo
  {author} {\bibfnamefont {D.}~\bibnamefont {Loffhagen}}, \bibinfo {author}
  {\bibfnamefont {S.}~\bibnamefont {Reuter}},\ and\ \bibinfo {author}
  {\bibfnamefont {K.}~\bibnamefont {Weltmann}},\ }\bibfield  {title} {\bibinfo
  {title} {Stability and excitation dynamics of an argon micro-scaled
  atmospheric pressure plasma jet},\ }\href@noop {} {\bibfield  {journal}
  {\bibinfo  {journal} {Plasma Sources Science and Technology}\ }\textbf
  {\bibinfo {volume} {24}},\ \bibinfo {pages} {065018} (\bibinfo {year}
  {2015})}\BibitemShut {NoStop}%
\bibitem [{\citenamefont {Fu}\ \emph {et~al.}(2020{\natexlab{b}})\citenamefont
  {Fu}, \citenamefont {Zheng}, \citenamefont {Wen}, \citenamefont {Zhang},
  \citenamefont {Fan},\ and\ \citenamefont {Verboncoeur}}]{fu2020similarity2}%
  \BibitemOpen
  \bibfield  {author} {\bibinfo {author} {\bibfnamefont {Y.}~\bibnamefont
  {Fu}}, \bibinfo {author} {\bibfnamefont {B.}~\bibnamefont {Zheng}}, \bibinfo
  {author} {\bibfnamefont {D.-Q.}\ \bibnamefont {Wen}}, \bibinfo {author}
  {\bibfnamefont {P.}~\bibnamefont {Zhang}}, \bibinfo {author} {\bibfnamefont
  {Q.~H.}\ \bibnamefont {Fan}},\ and\ \bibinfo {author} {\bibfnamefont {J.~P.}\
  \bibnamefont {Verboncoeur}},\ }\bibfield  {title} {\bibinfo {title}
  {Similarity law and frequency scaling in low-pressure capacitive radio
  frequency plasmas},\ }\href@noop {} {\bibfield  {journal} {\bibinfo
  {journal} {Applied Physics Letters}\ }\textbf {\bibinfo {volume} {117}},\
  \bibinfo {pages} {204101} (\bibinfo {year} {2020}{\natexlab{b}})}\BibitemShut
  {NoStop}%
\bibitem [{\citenamefont {Luque}\ and\ \citenamefont
  {Crosley}(1999)}]{luque1999lifbase}%
  \BibitemOpen
  \bibfield  {author} {\bibinfo {author} {\bibfnamefont {J.}~\bibnamefont
  {Luque}}\ and\ \bibinfo {author} {\bibfnamefont {D.~R.}\ \bibnamefont
  {Crosley}},\ }\bibfield  {title} {\bibinfo {title} {Lifbase: Database and
  spectral simulation program (version 1.5)},\ }\href@noop {} {\bibfield
  {journal} {\bibinfo  {journal} {SRI international report MP}\ }\textbf
  {\bibinfo {volume} {99}} (\bibinfo {year} {1999})}\BibitemShut {NoStop}%
\bibitem [{\citenamefont {Popov}(2011)}]{popov2011fast}%
  \BibitemOpen
  \bibfield  {author} {\bibinfo {author} {\bibfnamefont {N.}~\bibnamefont
  {Popov}},\ }\bibfield  {title} {\bibinfo {title} {Fast gas heating in a
  nitrogen--oxygen discharge plasma: I. kinetic mechanism},\ }\href@noop {}
  {\bibfield  {journal} {\bibinfo  {journal} {Journal of Physics D: Applied
  Physics}\ }\textbf {\bibinfo {volume} {44}},\ \bibinfo {pages} {285201}
  (\bibinfo {year} {2011})}\BibitemShut {NoStop}%
\bibitem [{\citenamefont {Lee}\ \emph {et~al.}(2017{\natexlab{c}})\citenamefont
  {Lee}, \citenamefont {Seo}, \citenamefont {Kwon}, \citenamefont {Kim},
  \citenamefont {Seong}, \citenamefont {Oh}, \citenamefont {Chung},
  \citenamefont {You},\ and\ \citenamefont {Shin}}]{lee2017evolution}%
  \BibitemOpen
  \bibfield  {author} {\bibinfo {author} {\bibfnamefont {H.-C.}\ \bibnamefont
  {Lee}}, \bibinfo {author} {\bibfnamefont {B.}~\bibnamefont {Seo}}, \bibinfo
  {author} {\bibfnamefont {D.-C.}\ \bibnamefont {Kwon}}, \bibinfo {author}
  {\bibfnamefont {J.}~\bibnamefont {Kim}}, \bibinfo {author} {\bibfnamefont
  {D.}~\bibnamefont {Seong}}, \bibinfo {author} {\bibfnamefont
  {S.}~\bibnamefont {Oh}}, \bibinfo {author} {\bibfnamefont {C.-W.}\
  \bibnamefont {Chung}}, \bibinfo {author} {\bibfnamefont {K.}~\bibnamefont
  {You}},\ and\ \bibinfo {author} {\bibfnamefont {C.}~\bibnamefont {Shin}},\
  }\bibfield  {title} {\bibinfo {title} {Evolution of electron temperature in
  inductively coupled plasma},\ }\href@noop {} {\bibfield  {journal} {\bibinfo
  {journal} {Applied Physics Letters}\ }\textbf {\bibinfo {volume} {110}},\
  \bibinfo {pages} {014106} (\bibinfo {year} {2017}{\natexlab{c}})}\BibitemShut
  {NoStop}%
\bibitem [{\citenamefont {Laux}\ \emph {et~al.}(2003)\citenamefont {Laux},
  \citenamefont {Spence}, \citenamefont {Kruger},\ and\ \citenamefont
  {Zare}}]{laux2003optical}%
  \BibitemOpen
  \bibfield  {author} {\bibinfo {author} {\bibfnamefont {C.~O.}\ \bibnamefont
  {Laux}}, \bibinfo {author} {\bibfnamefont {T.}~\bibnamefont {Spence}},
  \bibinfo {author} {\bibfnamefont {C.}~\bibnamefont {Kruger}},\ and\ \bibinfo
  {author} {\bibfnamefont {R.}~\bibnamefont {Zare}},\ }\bibfield  {title}
  {\bibinfo {title} {Optical diagnostics of atmospheric pressure air plasmas},\
  }\href@noop {} {\bibfield  {journal} {\bibinfo  {journal} {Plasma Sources
  Science and Technology}\ }\textbf {\bibinfo {volume} {12}},\ \bibinfo {pages}
  {125} (\bibinfo {year} {2003})}\BibitemShut {NoStop}%
\bibitem [{\citenamefont {Zheng-De}\ and\ \citenamefont
  {Yi-Kang}(2002)}]{zheng2002molecular}%
  \BibitemOpen
  \bibfield  {author} {\bibinfo {author} {\bibfnamefont {K.}~\bibnamefont
  {Zheng-De}}\ and\ \bibinfo {author} {\bibfnamefont {P.}~\bibnamefont
  {Yi-Kang}},\ }\bibfield  {title} {\bibinfo {title} {Molecular nitrogen
  vibrational temperature in an inductively coupled plasma},\ }\href@noop {}
  {\bibfield  {journal} {\bibinfo  {journal} {Chinese physics letters}\
  }\textbf {\bibinfo {volume} {19}},\ \bibinfo {pages} {211} (\bibinfo {year}
  {2002})}\BibitemShut {NoStop}%
\bibitem [{\citenamefont {Lieberman}\ and\ \citenamefont
  {Lichtenberg}(2005)}]{lieberman2005principles}%
  \BibitemOpen
  \bibfield  {author} {\bibinfo {author} {\bibfnamefont {M.~A.}\ \bibnamefont
  {Lieberman}}\ and\ \bibinfo {author} {\bibfnamefont {A.~J.}\ \bibnamefont
  {Lichtenberg}},\ }\href@noop {} {\emph {\bibinfo {title} {Principles of
  plasma discharges and materials processing}}}\ (\bibinfo  {publisher} {John
  Wiley \& Sons},\ \bibinfo {year} {2005})\BibitemShut {NoStop}%
\bibitem [{\citenamefont {Bruggeman}\ \emph {et~al.}(2014)\citenamefont
  {Bruggeman}, \citenamefont {Sadeghi}, \citenamefont {Schram},\ and\
  \citenamefont {Linss}}]{bruggeman2014gas}%
  \BibitemOpen
  \bibfield  {author} {\bibinfo {author} {\bibfnamefont {P.}~\bibnamefont
  {Bruggeman}}, \bibinfo {author} {\bibfnamefont {N.}~\bibnamefont {Sadeghi}},
  \bibinfo {author} {\bibfnamefont {D.}~\bibnamefont {Schram}},\ and\ \bibinfo
  {author} {\bibfnamefont {V.}~\bibnamefont {Linss}},\ }\bibfield  {title}
  {\bibinfo {title} {Gas temperature determination from rotational lines in
  non-equilibrium plasmas: a review},\ }\href@noop {} {\bibfield  {journal}
  {\bibinfo  {journal} {Plasma Sources Science and Technology}\ }\textbf
  {\bibinfo {volume} {23}},\ \bibinfo {pages} {023001} (\bibinfo {year}
  {2014})}\BibitemShut {NoStop}%
\end{thebibliography}

%apsrev4-2.bst 2019-01-14 (MD) hand-edited version of apsrev4-1.bst
%Control: key (0)
%Control: author (8) initials jnrlst
%Control: editor formatted (1) identically to author
%Control: production of article title (0) allowed
%Control: page (0) single
%Control: year (1) truncated
%Control: production of eprint (0) enabled
\providecommand{\noopsort}[1]{}\providecommand{\singleletter}[1]{#1}%

\end{document}